\documentclass[journal,twocolumn]{IEEEtran}

\usepackage{graphicx}
\usepackage{import}
\usepackage{dcolumn}
\usepackage{bm}
\usepackage{hyperref} 
\usepackage[T1]{fontenc}
\usepackage{siunitx}
\begin{document}

\title{Rapid readout of terahertz orbital angular momentum beams using atom-based imaging}


\author{\IEEEauthorblockN{Lucy A. Downes*, Daniel J. Whiting, Charles S. Adams, Kevin J. Weatherill}
   \\ \IEEEauthorblockA{Department of Physics, Durham University, DH1 3LE
    \\\*lucy.downes@durham.ac.uk}}





\maketitle

\begin{abstract}

We demonstrate the rapid readout of terahertz (THz) orbital angular momentum (OAM) beams using an atomic-vapour based imaging technique. OAM modes with both azimuthal and radial indices are created using phase-only transmission plates. The beams undergo terahertz to optical conversion in an atomic vapour, before being imaged in the far field using an optical CCD camera. In addition to the spatial intensity profile, we also observe the self-interferogram of the beams by imaging through a tilted lens, allowing the sign and magnitude of the azimuthal index to be read out directly. Using this technique, we can reliably read out the OAM mode of low-intensity beams with high fidelity in 10\,\si{\milli\second}. Such a demonstration is expected to have far-reaching consequences for proposed applications of terahertz OAM beams in communications and microscopy.
 
\end{abstract}


\section{Introduction}
The orbital angular momentum (OAM) of light has been of fundamental interest in optics research for the past thirty years \cite{Allen:92,Yao11} and OAM beams have found many applications, notably in free-space information transfer \cite{Gibson:04, Wang22} and encryption \cite{Fang2020}. 
Recently, there has been growing interest in OAM beams outside of the optical frequency range. For example, Bragg mirrors have been used to create OAM X-ray beams from a free-electron laser \cite{Huang2021} and microwave OAM beams have been applied to radar imaging techniques \cite{Yan16, Liu2020}. There has also been considerable interest in the creation, manipulation and detection of OAM beams in the THz frequency range \cite{Wang_2020} for applications as diverse as short-range communications \cite{Troha:21}, super resolution imaging \cite{Miyamoto:2014} and astronomy \cite{Harwit_2003}. 
The creation of THz OAM beams has been demonstrated using numerous methods including spiral phase-plates \cite{Turnbull:96,Miyamoto:14}, helical axicons \cite{Wei:15}, V-shaped antenna structures \cite{He:13}, polarizing optical elements \cite{Imai:14}, THz spatial light modulators \cite{Xie:13} and soft-aperture difference frequency generation \cite{Miyamoto:19}.
However, the detection and readout of these beams remains challenging, largely due to the lack of fast and sensitive terahertz detectors \cite{Mittleman:18}. 

In this work we demonstrate the rapid (10~ms) readout of terahertz OAM beam modes using a recently-developed, atom-based THz imaging technology \cite{Wade17, Downes:20}, and show that we can distinguish between beams of varying azimuthal and radial phase index. This work offers a way forward in realising the demanding applications of THz OAM beams. 

\section{Experiment}
\begin{figure}[ht]
    \centering
    \includegraphics[width=0.95\linewidth]{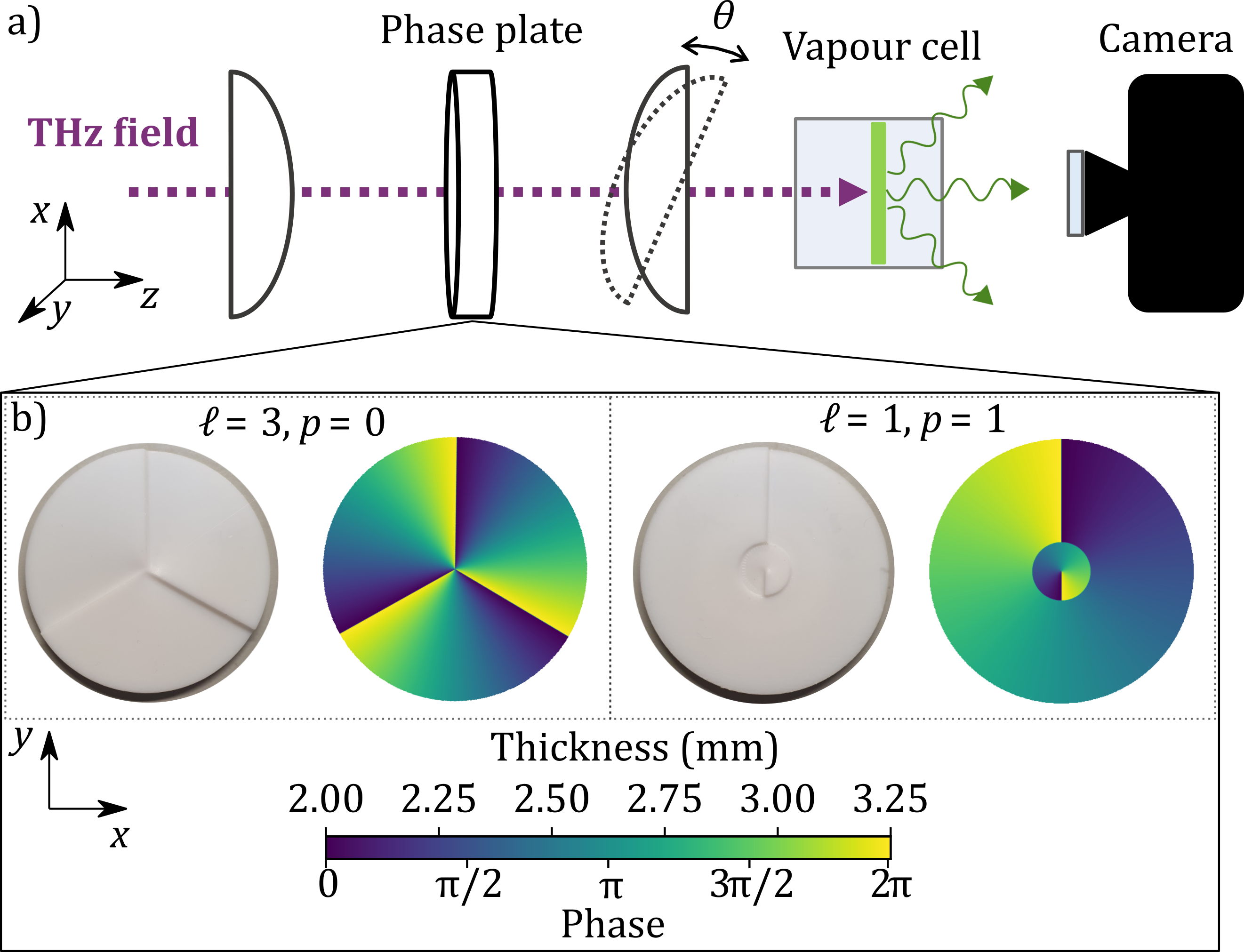}
    \caption[Experimental setup and phase plate design]{\textbf{Experimental setup and phase plate design} a) The phase plates are placed at the centre of a collimated THz beam (purple dotted line) which is focused onto an atomic vapour using a second Teflon lens. This lens can be tilted with respect to the phase plate at an angle $\theta$, where $\theta = 0\si{\degree}$ corresponds to the lens being parallel to the phase plate. The resulting fluorescence from the atomic vapour is then imaged using an optical camera. b) Photographs and thickness/phase profiles of the 50\,\si{\milli\metre} diameter Teflon plates used to create THz OAM beams, designed using the method described in \cite{Fasbender:18}. The design frequency is 550\,\si{\giga\hertz}, equal to that of the THz field used in the atomic imaging system. Note that the constant minimum thickness of 2\,\si{\milli\metre} is a result of manufacturing constraints and does not alter the phase profile.}
    \label{fig:SPP_sim}
\end{figure}

The experimental setup is shown in Fig.~\ref{fig:SPP_sim}a). The THz field is a 20\,\si{\micro\watt}, 0.55\,\si{\tera\hertz} free space beam produced by an amplifier multiplier chain and diagonal horn antenna. 
The beam is predominantly a TEM$_{00}$ mode with 84\% Gaussian mode content \cite{VDI_Horn} and is collimated using a Teflon lens with focal length $f = 75\,\si{\milli\metre}$ to give a $1/e^2$ radius of $w_0 = 5.55\,\si{\milli\metre}$. Teflon phase plates, manufactured by CNC machining, are then placed in the centre of the THz beam, at a position approximately $75\,\si{\milli\metre}$ from the first lens. The resulting beam passes though a second $f = 75\,\si{\milli\metre}$ Teflon lens and is imaged in the focal plane. 
The final Teflon lens can be positioned parallel to the phase plate (solid lines, Fig.~\ref{fig:SPP_sim}a)) or tilted with respect to the phase plate (dashed lines, Fig.~\ref{fig:SPP_sim}a)). We refer to the image observed in the case of the parallel lens as the intensity pattern and the image produced by the tilted lens as the auto-interference pattern \cite{Chopinaud:18}. The theory behind the differences in the structure of these patterns is described in detail in \cite{Vaity:13}. Photographs of example phase plates are shown in Fig.~\ref{fig:SPP_sim}b) along with plots showing the thickness of the Teflon and the resulting phase profile of the plates. Both azimuthal phase, $\ell$, and radial phase, $p$, can be applied.

The method used to image the THz field is the same as described in \cite{Downes:20} and utilises the concept of THz-to-optical conversion by an atomic vapour \cite{Wade17}. Briefly, three resonant infrared lasers, stabilised to atomic transitions \cite{Pearman_2002,Carr:12,Carr12b}, are used to excite caesium atoms to a high-lying energy state, a so-called Rydberg state. 
All Rydberg transition frequencies are calculated using the ARC Python package \cite{Sibalic17}. 
The laser beams are shaped such that, at the position of the glass cell containing Cs vapour, they overlap and form a light sheet of approximately $10\,\si{\milli\metre} \times 10\,\si{\milli\metre} \times 100\,\si{\micro\metre}$ in the $x, y$ and $z$ directions respectively. 
The Rydberg atoms decay back to the electronic ground state through the emission of visible fluorescence, the wavelength of which depends on the pathway taken. Once excited, interaction with a resonant THz field transfers atomic population from the initial Rydberg state to another nearby Rydberg state, from which there is one dominant decay pathway leading to the emission of green (535\,\si{\nano\metre}) photons. Therefore, by looking at the intensity of the 535\,\si{\nano\metre} fluorescence emitted by the vapour we can infer the spatial distribution of the incident THz field. 
The THz beam propagates perpendicular to the IR lasers used to create the light sheet and the atomic fluorescence is imaged from the opposite side of the vapour cell to that on which the THz is incident, as shown in Fig.~\ref{fig:SPP_sim}a). Throughout this work we use an Andor iXon EMCCD to image the optical fluorescence from the vapour with a $(535 \pm 6)\,\si{\nano\metre}$ bandpass filter to eliminate unwanted background light. All images are taken using an exposure time of 100\,\si{\milli\second} unless otherwise stated, and an image of the vapour in the absence of the THz field is subtracted to remove background fluorescence and increase contrast.

\section{Results and Discussion}

\begin{figure}
    \centering
    \includegraphics[width = 0.8\linewidth]{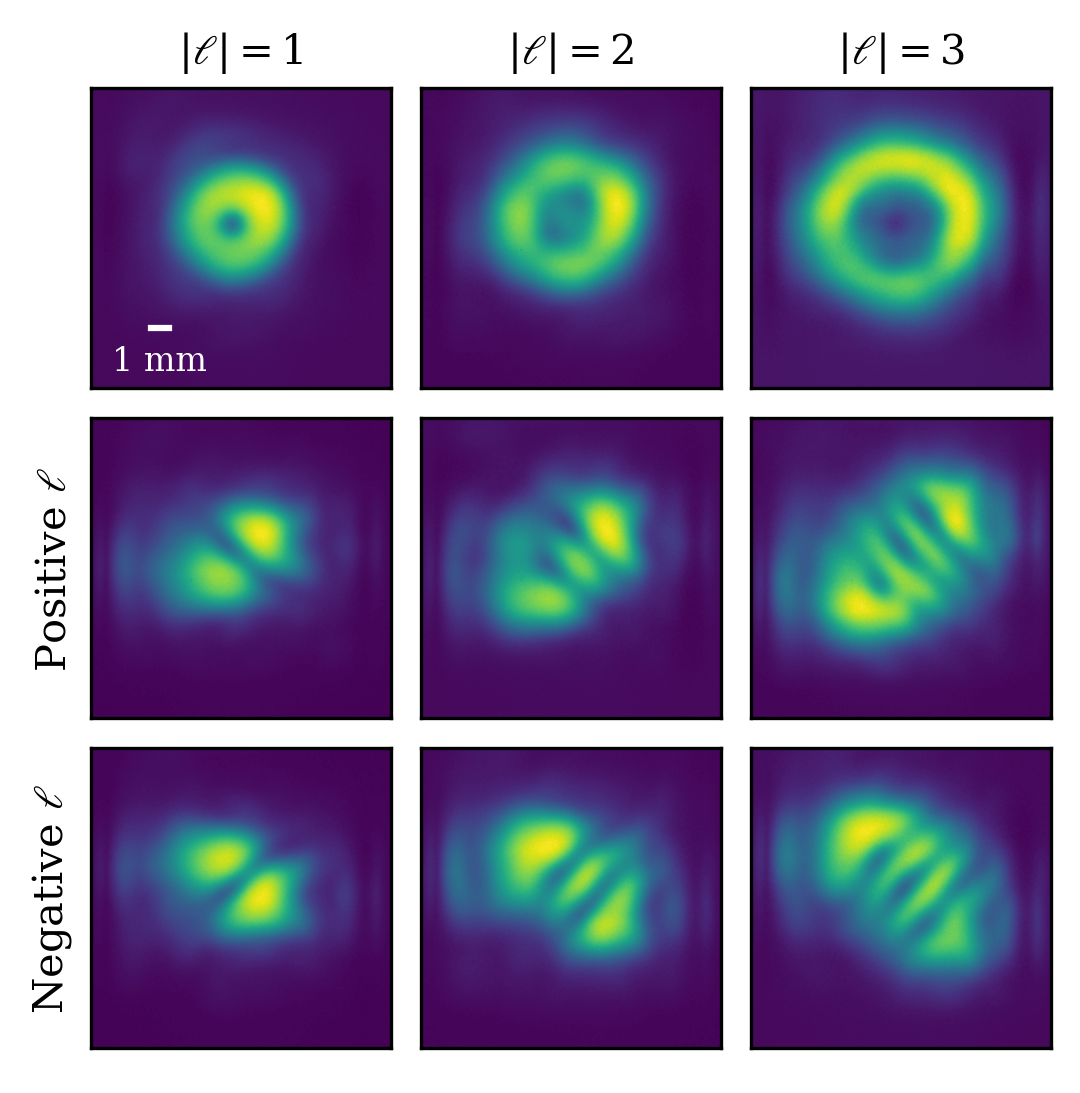}
    \caption[Intensity patterns of THz OAM beams]{\textbf{Intensity and auto-interference patterns of THz OAM beams created using spiral phase plates.} Intensity patterns of beams with $p=0,\: |\ell|=1,2,3$. The top row shows the images produced when the second lens is parallel to the phase plate ($\theta = 0\si{\degree}$) while the second and third rows show the resulting intensity pattern for a lens tilt of $\theta\approx 30\si{\degree}$. Since the donut-like patterns are identical for $\ell = -\ell$, only the images for positive $l$ are shown here. The second and third rows show the auto-interference patterns for positive and negative values of $\ell$ respectively.}
    \label{fig:all_beams}
\end{figure}

Images of the intensity patterns produced for beams with $p=0,\,\ell=1,2,3$ are shown in Fig.~\ref{fig:all_beams}. The top row shows the beam intensity imaged with the lens parallel to the phase plate ($\theta = 0\si{\degree}$) for increasing values of $\ell$. 
As expected the intensity patterns for beams with $p=0, \ell\neq 0$ have a `donut-like' structure; a central vortex surrounded by a single ring, the radius of which is determined by the value of $\ell$.
Since these intensity patterns are identical for $\ell = -\ell$ only the images for beams with positive values of $\ell$ are shown. 
The lower two rows show the auto-interference patterns resulting from a lens tilt of $\theta\approx 30\si{\degree}$. 
These patterns not only display a number of dark fringes equal to the value of $\ell$, but also exhibit a $45\si{\degree}$ rotation in the imaging plane. The direction of this rotation depends on the sign of $\ell$, allowing this to be determined from the pattern. This rotation is not affected by the direction of tilt of the final lens which remained unchanged between rows two and three.

\begin{figure}
    \centering
    \includegraphics[width = 0.9\linewidth]{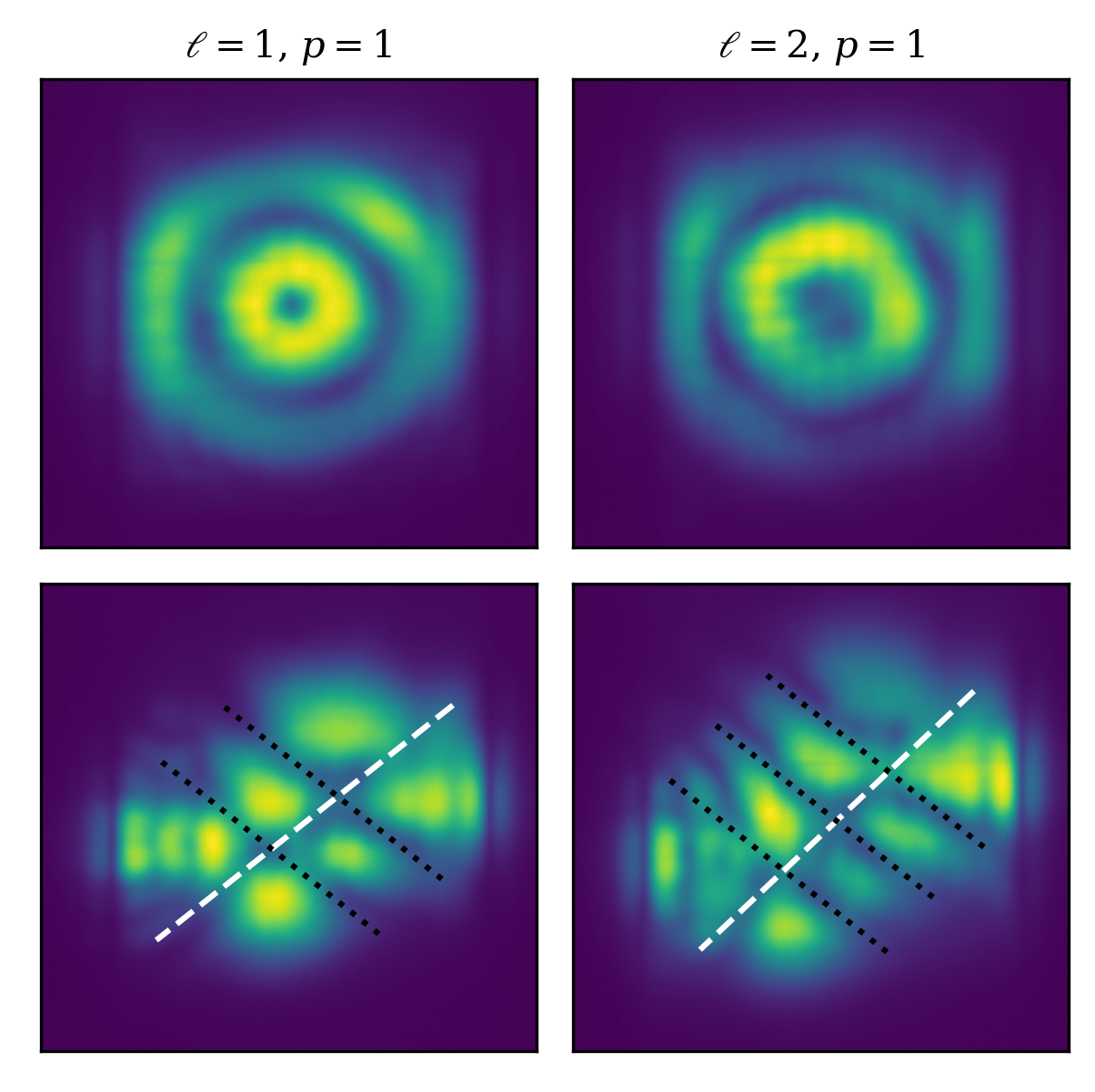}
    \caption[Intensity patterns for beams with $\ell\neq 0$ and $p\neq 0$]{\textbf{Intensity and auto-interference patterns for beams with $\mathbf{\boldsymbol{\mathit{\ell}}\neq 0}$ and $\mathbf{\boldsymbol{\mathit{p}}\neq 0}$.} Intensity (\textit{top}) and auto-interference patterns (\textit{bottom}) for beams with $p = 1,\:\ell= 1,2$ (\textit{left and right columns respectively}). The white dashed and black dotted lines have been added to the auto-interference patterns to highlight the dark fringes from which we can extract the values of $p$ and $p+|\ell|$. Only the auto-interference patterns for positive $\ell$ values are shown here, those for negative $\ell$ values would be tilted in the opposite direction with respect to the vertical.}
    \label{fig:lp_beams}
\end{figure}

Figure~\ref{fig:lp_beams} shows the intensity (top row) and auto-interference patterns (bottom row) for beams with $p = 1,\:\ell=1,2$. The intensity patterns of beams with both $\ell\neq 0$ and $p\neq 0$ have a central vortex surrounded by a number of rings equal to $p+1$. The radius of the rings is again determined by the value of $\ell$. 
Their auto-interference patterns exhibit orthogonal dark fringes, where the number of fringes corresponds to the values of $p$ and $p+|\ell|$ (shown in the second row of Fig.~\ref{fig:lp_beams} by the white dashed and black dotted lines respectively).
Again the rotation of these fringe patterns depends solely on the sign of the index $\ell$. 
As $p$ increases the beams produced become larger and no longer fit within our $1\,\si{\centi\metre}^2$ imaging area. As a result we see interference patterns at the edges of the image where the THz field is reflected from the walls of the glass cell containing the atomic vapour. These interference effects can be seen as vertical stripes at the edges of the images in Fig.~\ref{fig:lp_beams}, and is the reason why we were unable to image beams with $p>1$ clearly.

\begin{figure}
    \centering
    \includegraphics[width = 0.9\linewidth]{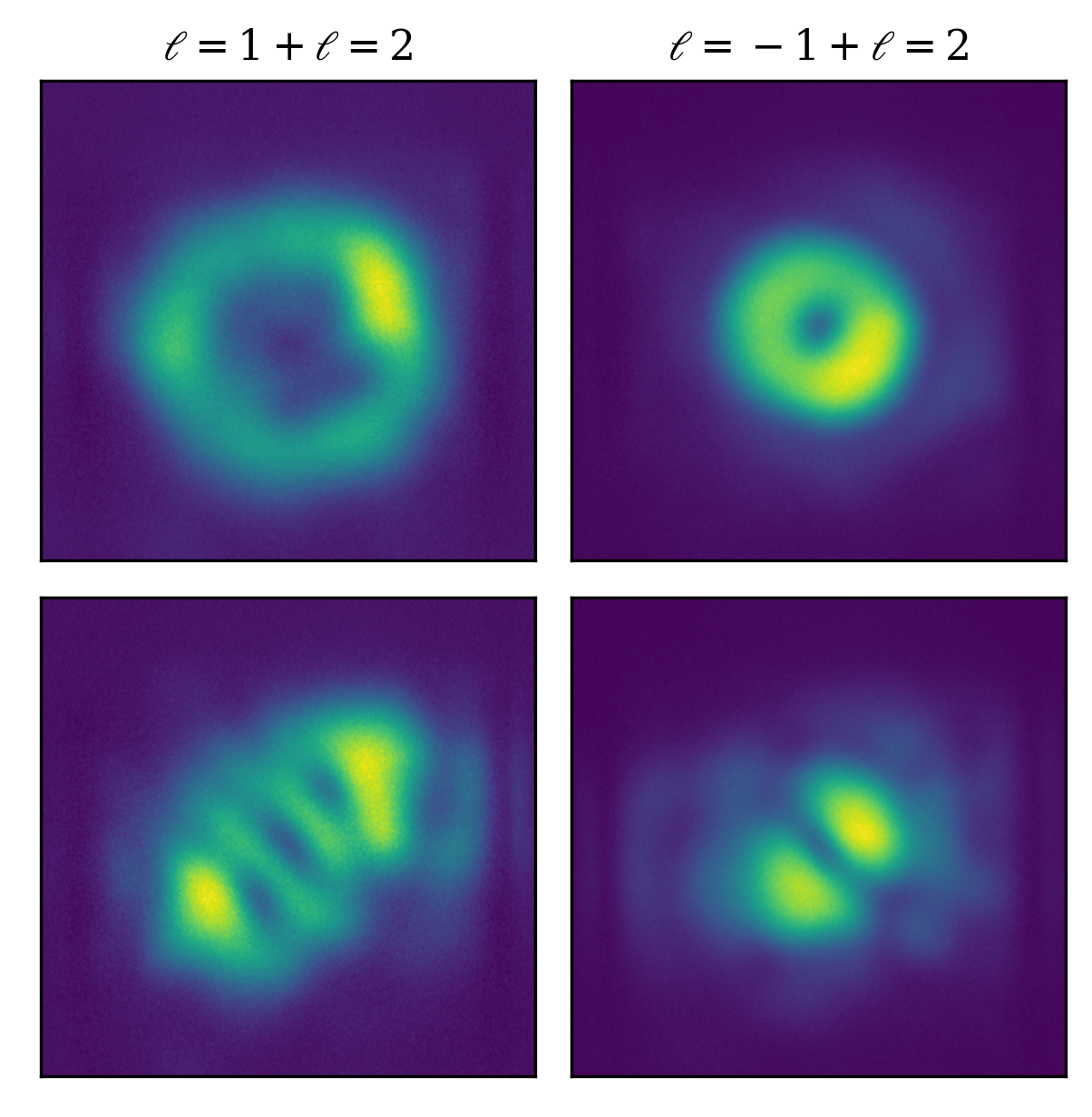}
    \caption[Combining phase plates]{\textbf{Combinations of phase plates.} Intensity (\textit{top}) and auto-interference patterns (\textit{bottom}) for OAM beams produced by combining phase plates with $p=0,\:\ell=1$ and $p=0,\:\ell=2$. In the left hand column the plates were placed in the same direction, hence the resulting beam has $\ell=3$. In the second column the $\ell=1$ phase plate was reversed resulting in the final beam having $\ell=1$.}
    \label{fig:sum_plates}
\end{figure}

By using phase plates made from a material with a low absorption coefficient in this frequency range such as Teflon ($\alpha = 0.2\,\si{\centi\metre}^{-1}$ at 550\,\si{\giga\hertz} \cite{Winnewisser:98}) we achieve over 90\% transmission, even for the thickest plates used in this work ($\approx 5\,\si{\milli\metre}$). 
Because of these low losses we can combine phase plates additively to form beams with the sum of the indices of the plates used. For example placing the $\ell=1,\:p=0$ and the $\ell=2,\:p=0$ plates sequentially in the beam creates an intensity pattern equivalent to the $\ell=3,\:p=0$ case. This is shown in the left hand column of Figure~\ref{fig:sum_plates}. Similarly reversing the direction of the $\ell=1,\:p=0$ plate creates a beam with intensity pattern equivalent to the $\ell=1,\:p=0$ case, as shown on the right hand side of Figure~\ref{fig:sum_plates}. In both cases the images created using multiple plates are not as clear as those created using a single plate of equivalent topological charge. This is due to unwanted aberrations induced by each plate. 

From Figures~\ref{fig:all_beams} and \ref{fig:lp_beams} we observe that the radius of the rings scales approximately linearly with the value of $|\ell|$. Measuring the radius of maximum intensity ($r_{\ell}$) for the intensity patterns produced by the spiral phase plates for the $p=0$ case (top row, Fig.~\ref{fig:all_beams}) we find that they scale as $r_{\ell=2} = 1.92\times r_{\ell=1}$ and $r_{\ell=3} = 2.85\times r_{\ell=1}$. 
For a pure Laguerre-Gauss mode we would expect the radius of maximum intensity to scale as $\sqrt{|\ell|}$  \cite{Padgett:15} however, a linear scaling has previously been observed in optical vortices created using spiral phase plates \cite{Kotlyar:06, Saitoh:12, Harm:15}. This discrepancy in scaling is because the beams created by the phase plates are not pure LG modes but are instead a superposition of modes with different $\ell$ and $p$, meaning that the resulting beams do not share the same scaling properties. 

\begin{figure}[hbt]
    \centering
    \includegraphics[width = 0.9\linewidth]{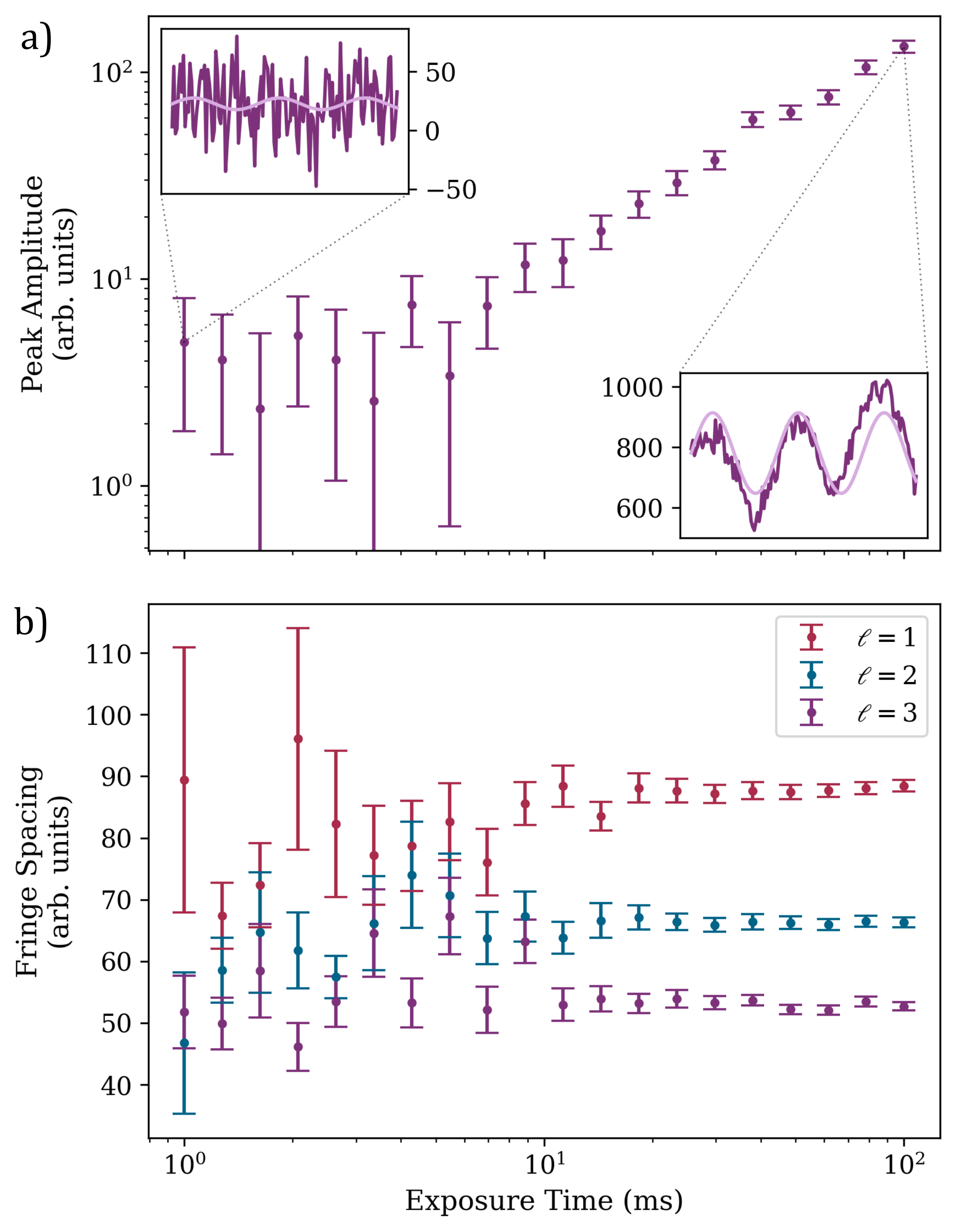}
    \caption[Fidelity]{\textbf{Fringe visibility at reduced exposure times.} a) Fitted fringe amplitude as a function of exposure time for the $p=0,\:\ell=3$ auto-interference pattern. Below $10~\si{\milli\second}$ the fitted amplitude is lower than its associated uncertainty so we say that the fringes are indistinguishable at this point. Insets show data (dark purple) and corresponding fits (light purple) for the shortest (\textit{top left}) and longest (\textit{bottom right}) exposures. b) Fitted fringe spacing as a function of exposure time for $p=0,\:\ell=1,2,3$. Again below $10~\si{\milli\second}$ the fitted fringe spacing is not consistent, and hence the value of $\ell$ cannot be determined.}
    \label{fig:fidelity}
\end{figure}

To demonstrate how the fast readout capabilities of the imaging technique presented in \cite{Downes:20} could be applied to OAM beams, we image auto-interference patterns as the exposure time is reduced. 
For each image a cross-section is taken along the diagonal and a sine function fitted to the fringe pattern. Initially the fringe spacing and position are fixed in the fitting, the only free parameters are the amplitude of the fringes and the background level. This is shown in Fig.~\ref{fig:fidelity}a) for beams with $p=0, \ell=3$.
From this analysis it can be seen that the fringes are distinguishable at exposure times greater than $10~\si{\milli\second}$. At shorter exposure times the fitted amplitude of the fringes is less than the associated error and hence the fringes cannot be distinguished from the noise. 
Fig.~\ref{fig:fidelity}b) shows the results of letting both the fringe amplitude and spacing vary for cross-sections of the $p=0,\:\ell=1,2,3$ auto-interference patterns. Again, at exposure times greater than $10~\si{\milli\second}$ the fitting can reliably distinguish between the different orders of $\ell$, shown by the consistency of the fitted spacing. 
We note that the limit of $10~\si{\milli\second}$ presented here is specific to the camera used and may be reduced by using a camera better suited to shorter exposures.

While phase plates present an easy way to create and manipulate arbitrary OAM beams in the THz frequency range, in order to use these beams for information transfer a method of quickly switching between values of $\ell$ and $p$ would be needed to achieve reasonable data transfer rates. 
Fast switching could be realised by using a spatial light modulator (SLM) \cite{Stantchev2020} to rapidly modify the helicity of the THz beam, or by using a programmable reflectarray antenna \cite{Soleimani:2021}. 
Combining these techniques with the imaging method presented here would enable fast information transfer in the THz regime. 
Furthermore, the use of an atomic detection system allows for the calibrated electric field measurement of such beams \cite{Wade17,Chen:22}.

\section{Conclusion}
We have demonstrated high-speed readout of THz OAM beams created using Teflon phase plates with both varying $\ell$ and $p$ indices. We have also demonstrated that the sign of the phase can be determined by imaging the beam through a tilted lens, and that this can be done at exposure times down to 10\,\si{\milli\second}. We note that these phase plates do not create pure Laguerre-Gauss modes as demonstrated by the fact that the vortex radius scales with $\ell$ and not with $\sqrt{\ell}$ as would be expected for a pure mode. This high-speed method of determining the characteristics of THz OAM beams could enable their use in free-space data transfer. \\

\noindent\textbf{Acknowledgements:} We thank Stephen Lishman, Lee Bainbridge and Andrew Davies for the manufacture of the phase plates, Neal Radwell, Andrew Forbes and Ifan Hughes for fruitful discussions and Mike Tarbutt for the loan of equipment.

\noindent\textbf{Funding:} Engineering and Physical Sciences Research Council (EP/R002061/1, EP/S015973/1). European Commission Quantum Technology Flagship project macQsimal (820393)

\bibliography{OAMbib}
\bibliographystyle{IEEEtran}

\end{document}